\documentclass
{mn2e}
\usepackage{epsfig}
\usepackage{amsmath, amssymb,bm}
\usepackage{color}

\def \be{\begin{equation}}
\def \ee{\end{equation}}
\def \msun{\rm M_{\odot}}
\def \me{{\dot M_{\rm Edd}}}
\def \le{{L_{\rm Edd}}}
 
\begin{document}
\title[Dwarf Galaxies and the Black--Hole Scaling Relations]{Dwarf Galaxies and the Black--Hole Scaling Relations}

\author[Andrew King \& Rebecca Nealon] 
{\parbox{5in}{Andrew King$^{1, 2, 3}$ 
\& Rebecca Nealon$^{1,4,5}$}
\vspace{0.1in} \\
$^1$ Department of Physics \& Astronomy, University
of Leicester, Leicester LE1 7RH UK\\ 
$^2$ Astronomical Institute Anton Pannekoek, University of Amsterdam, Science Park 904, NL-1098 XH Amsterdam, The Netherlands \\
$^{3}$ Leiden Observatory, Leiden University, Niels Bohrweg 2, NL-2333 CA Leiden, Netherlands \\
$^{4}$ Centre for Exoplanets and Habitability, University of Warwick, Coventry CV4 7AL, UK \\
$^{5}$ Department of Physics, University of Warwick, Coventry CV4 7AL, UK
}


\maketitle

\begin{abstract}
The sample of dwarf galaxies with measured central black hole masses $M$ and velocity dispersions $\sigma$ has recently doubled, and gives a close fit to the extrapolation of the $M \propto \sigma$ relation for more massive galaxies. We argue that this is difficult to reconcile with suggestions that the scaling relations between galaxies and their central black holes are simply a statistical consequence of assembly through repeated mergers. This predicts black hole masses significantly larger than those observed in dwarf galaxies unless the initial distribution of uncorrelated seed black hole and stellar masses is confined to much smaller masses than earlier assumed. It also predicts a noticeable flattening of the $M \propto \sigma$ relation for dwarfs, to $M \propto \sigma^2$ compared with the observed $M \propto \sigma^4$. In contrast black hole feedback predicts that black hole masses tend towards a universal $M \propto \sigma^4$ relation in all galaxies, and correctly gives the properties of powerful outflows recently observed in dwarf galaxies. These considerations emphasize once again that the fundamental physical black-hole --- galaxy scaling relation is between $M$ and $\sigma$. The relation of $M$ to the bulge mass $M_b$ is acausal, and depends on the quite independent connection between $M_b$ and $\sigma$ set by stellar feedback.
\end{abstract}


\footnotetext[1]{E-mail: ark@astro.le.ac.uk}
\section{Introduction}
\label{intro}
It is now widely accepted that the centre of every medium-- or high--mass 
($\gtrsim 10^{10}\msun$) galaxy contains a supermassive black hole (SMBH). 
The hole mass $M$ is observed to scale with both the velocity dispersion 
$\sigma$ 
of the host galaxy spheroid (or bulge) and the bulge stellar mass $M_b$ as
\begin{equation}
M \propto \sigma^{\alpha},\,M \sim 10^{-3}M_b
\label{scaling}
\end{equation}
with $\alpha \sim 4$
(see e.g. Kormendy \& Ho 2013 for a review). These scalings give important 
constraints on how the SMBH and their host galaxies evolve. In this paper we 
argue that recent observations of the central black holes in dwarf galaxies 
distinguish sharply between two approaches to understanding the scaling 
relations.

One picture of these relations uses the fact that the SMBH binding energy 
$E_{\rm BH} = \eta Mc^2$ (where $\eta \sim 0.1$ is the accretion efficiency)
is typically $> 1000\times$ the binding energy $\sim 
f_gM_b\sigma^2$ of the bulge gas (where $f_g \sim 0.16$ is the gas fraction) of 
the host galaxy. (We use the term `bulge' to include pseudobulges also, where the 
velocities are dominated by ordered rotation. The 
distinction has no significance for either picture of the scaling relations, assuming
that the observed velocities dynamically specify the mass distributions.)

So it is plausible that the scaling relations (\ref{scaling}) may result from 
feedback via the powerful `UFO' (UltraFast Outflow)
winds observed from the accreting SMBH. These carry the Eddington 
momentum (i.e. $\dot M_w v_w \simeq \le/c$ 
(see King \& Pounds 2015 for a review). 
The shocks of the wind against the bulge gas 
cool rapidly (giving `momentum--driven' feedback) for SMBH masses $M$ less than
\begin{equation}
M_{\sigma}\simeq 3\times 
10^8\sigma_{200}^4\msun 
\label{msignum}
\end{equation}
(King, 2003; here $\sigma_{200} = \sigma/(200\,{\rm km\,s^{-1}}$)) and
push the surrounding gas into a thin shell which expands but cannot escape 
the galaxy. But at the mass (\ref{msignum}), shock cooling becomes 
ineffective and the wind now does expel the gas that would have fuelled 
any significant further SMBH growth, in an `energy--driven'
outflow (King, 2005). An expulsive Eddington wind is even more likely in dwarf
galaxies than in the larger ones for which the theory is well verified, since the 
ratio of the dynamical mass inflow rate $\dot M_{\rm dyn} \simeq 
f_g\sigma^3/G$ potentially driving accretion to the required Eddington accretion rate  $\me \propto \le \propto M \propto \sigma^4$ goes as ${\sigma}^{-1}\propto M^{-1/4}$. The predicted limiting mass 
(\ref{msignum}) is in good agreement with 
observations (see Section 2 below) 
of the SMBH mass as a function of $\sigma$. Observational selection
effects (see Batcheldor 2010) make it difficult to measure black hole masses 
significantly below this value for a given $\sigma$, in particular because of 
the need to resolve the 
SMBH sphere of influence (radius $\propto M\sigma^{-2}$), 
so observationally--determined SMBH
masses tend to lie close to the relation (\ref{msignum}).
The $M - M_b$ relation now follows `acausally' (cf Power et al. 2011)
since the observed Faber--Jackson (FJ) relation 
\be
L_b \propto \sigma^4,
\label{fj1}
\ee
where $L_b$ is the luminosity of the bulge stars (Faber \& Jackson 1976) 
and the assumption of a standard mass--to--light ratio $M_b \propto L_b$
for these stars gives a parallel relation $M_b \propto 
\sigma^4$ (cf Murray et al, 2005). This parallelism arises because 
the SMBH and the bulge stars each separately drive momentum--driven 
feedback, respectively via UFOs, and stellar winds and supernovae. These 
separately limit $M$ and $M_b$ to values which are each 
proportional to $\sigma^4$, but differ by a factor $\sim10^3$.  
Importantly, unlike 
the $M - \sigma$ relation, there is no physics in 
the connection between $M$ and $M_b$ -- that is, black holes do not set $M_b$. 

Both forms of feedback are present in dwarf galaxies, and in
particular vigorous AGN--driven winds are directly observed in them, as 
we discuss in Section 4 below. The feedback these produce depends only on the
current black hole mass, irrespective of the previous history of
SMBH growth, provided only that most of this mass was acquired by gas 
accretion. This is expected at least at low redshift from the Soltan (1982) relation.

A very different idea (Peng, 2007;  Jahnke \& Macci\`o, 2011) asserts 
that the scaling relations  (\ref{scaling})
are not a result of black hole feedback, and are instead
largely statistical. If the SMBH and galaxy spheroids satisfying these relations 
are built from mergers of large numbers of much smaller galaxies 
with uncorrelated stellar and black hole masses,  the central limit theorem implies
a linear 
relation $M \propto M_b$, with a dispersion tightening for larger $M, M_b$
because on average more mergers have taken place.
In practice, to improve the fit to the observed $M - M_b$ relation,  Jahnke
\& Macci\`o (2011) go beyond the pure merger picture by adding 
in the effects of star formation, black hole accretion, and the conversion to bulge 
mass of a fraction of the stars formed in the disc component of each halo.
They do not explicitly derive an $M - \sigma$ relation, but 
for high--mass galaxies this follows, since the
FJ relation gives $M_b \propto \sigma^4$, which then implies $M\propto 
\sigma^4$. In this assembly picture the normalizations of both the scaling 
relations (\ref{scaling}) 
are presumably fixed by the original uncorrelated mass
distributions of black holes in small galaxies before any mergers 
take place. 

These two pictures -- feedback or assembly -- predict very different
outcomes for low--mass galaxies. In the feedback picture all galaxies limit the
growth of their central black holes through the physics producing the
$M -\sigma$ relation, so we expect this relation to hold for dwarf galaxies,
and we expect to see energy--driven winds driving away the gas that would 
otherwise increase the black hole mass above $M_{\sigma}$. 
But in the assembly picture galaxies of sufficiently low mass
do not experience enough mergers to produce a tight relation 
between $M$ and $M_b$ (cf Jahnke \& Macci\`o, 2011, Fig. 4). Further, 
we will see that this picture predicts a significant flattening 
($M_{\sigma} \propto \sigma^2$) in the  $M - \sigma$ relation
at low galaxy masses, contrary to observation. As this implies SMBH which are
{\it more} massive than expected from a simple extrapolation of the $M - \sigma$ relation for higher--mass galaxies, the fact that observations do not
seem to find them is significant.

These distinctions between feedback and assembly mean that observations of 
dwarf galaxies potentially give clean tests of whether either theory offers viable 
explanations of the scaling relations. Recent papers report observations of two 
types bearing directly on this question, and we discuss these in the rest of this 
paper.

\section{The $M - \sigma$ relation for dwarf galaxies}

Baldassare et al. (2020) used the Keck Echellete Spectrograph and Imager to 
measure stellar velocity dispersions for eight active dwarf galaxies ($M_b < 
3\times 10^9\msun$) with virial black hole masses. This
increases from 7 to 15 the number of dwarf galaxies 
which have measurements of both the black--hole mass $M$ and the velocity 
dispersion $\sigma$. This combined sample fits tightly on to the extrapolation of 
the $M - \sigma$ relation to low black--hole masses $M \lesssim 10^6\msun$ 
(Baldassare et al. 2020, Fig. 3). In addition Davis et al (2020) used sub--parsec
resolution ALMA observations to find a further dwarf galaxy (NGC 404)
lying on the $M - \sigma$ relation, with $M \simeq 5\times 10^5\msun$ and 
$M_b \sim 10^9\msun$.  Here both the observed molecular gas and the stellar
kinematics independently require this same black hole mass.

These results discriminate sharply between feedback and assembly.
In the feedback picture the physics producing the $M - \sigma$
relation holds for all galaxy masses, so the extrapolation to lower masses is 
unproblematic. But this same extrapolation runs strongly against the assembly 
theory. First, this produces a large scatter in black hole masses at low galaxy 
masses. Fig. 4 of Jahnke \& Macci\`o (2011) predicts a significant population of 
SMBH with 
masses $M \gtrsim 10^7\msun$ at stellar masses $M_* \lesssim 10^9\msun$. 
These black hole masses are considerably higher than those observed. Since 
they would have larger spheres of influence, in which stars move with higher 
velocities, it is unlikely that they have been missed because of selection effects. 
Evidently this problem arises because the maximum masses of the initial seed
black holes allow many of 
them to exceed observed black hole masses after only a few 
mergers. So one might try to alleviate the problem by reducing the initial black--
hole mass scatter below the $10^4$ range adopted by  Jahnke \& Macci\`o 
(2011). Since the predicted 
low--redshift scatter scales roughly as $\surd{N}$, and is about an order
of magnitude above observations, this reduces the required initial scatter 
in black hole mass to a factor $\lesssim 100$.

This already makes the assembly picture considerably less attractive, but 
a second problem for it appears in deriving the $M - \sigma$ 
relation at low galaxy masses. Instead of the $M_b \propto \sigma^4$ relation 
which follows from the FJ relation for massive galaxies, 
galaxies with velocity dispersions $\sigma \lesssim 100\,{\rm\, 
km\,s^{-1}}$ instead obey
\be
M_b \propto L_b \propto \sigma^2
\label{fjdwarf}
\ee
(Kourkchi et al., 2012: see also Davies et al. 1983; Held et al. 1992; and Matkovi\'c 
\& Guzm\'an, 2005, de Rijcke et al, 2005). Assuming continuity  between the 
two relations (\ref{fj1}, \ref{fjdwarf}) at $\sigma \simeq 100\,{\rm\, 
km\,s^{-1}}$ inevitably means that the flatter $\sigma^2$ relation gives an 
$M_b$ value
$4\times$ larger than given by the $\sigma^4$ relation at $\sigma = 50\,{\rm\, 
km\,s^{-1}}$. Taking $M \simeq 10^{-3}M_b$ (H\"aring
\& Rix, 2004) gives 
\begin{equation}
M_b \simeq 2\times 10^9\sigma^2_{50} \msun
\label{fj}
\end{equation}
Here $\sigma_{50}$ is the galaxy velocity dispersion $\sigma$ in units of 
$50\,{\rm\, km\,s^{-1}}$.
This is flatter than the $M_b \propto \sigma^4$ Faber--Jackson relation
found for large galaxies, and implies that the stellar 
components of dwarf galaxies all have roughly similar radii $R_b$. 
Approximating dwarfs as isothermal spheres, i.e.
\begin{equation} 
R_b \simeq \frac{GM_b}{2\sigma^2} 
\label{iso}
\end{equation}
gives $R_b$ of order 1 kpc
largely independently of $M_b$ or $\sigma$ -- we will find a best--fit value
\begin{equation}
R_b \simeq 2.3 \pm 1.1\,{\rm kpc}
\label{r}
\end{equation}
(cf Fig. 2). Inspection of Fig. 2 of Manzano--King et al. (2019) confirms that this is a 
reasonable approximation for the sizes of the hosts in their dwarf AGN sample
(see the discussion in Section 3 below). Adopting (\ref{r}) avoids the need 
to assign mass--to--light ratios for these small galaxies.
The origin of the near--constant radius (\ref{r}) is unclear, but cosmological 
simulations 
do find this effect at low masses (Furlong et al., 2017; but see also Ludlow et al., 
2019). A possible physical cause may relate to the fact that at gas
temperatures $\sim 10^4\, {\rm K}$ typical of the warm ISM,
the Jeans length is of order 1 kpc.

Since assembly can only ever 
give a linear $M - M_b$ relation, it predicts an $M - 
\sigma$ relation flattening to  
\begin{equation}
M_{\sigma} \simeq 2\times 10^6
\sigma_{50}^2\msun 
\label{msig1}
\end{equation}
for $\sigma \lesssim 100\,{\rm\, km\,s^{-1}}$. 

Kormendy \& Ho (2012)
find a larger normalization $M \simeq 5\times 10^{-3}M_b$ for the $M - M_b$
relation than H\"aring 
\& Rix (2001), so this relation would become 
\be
M_{\sigma} \simeq 1\times 10^7\sigma_{50}^2\msun 
\label{msig2}
\ee
in this case. We note that from (\ref{iso}) that this normalization
implies rather large radii for dwarfs
compared with the sizes seen in Fig. 2 of Manzano--King et al., (2019). 

We plot the two relations (\ref{msig1}, \ref{msig2}) in Fig. 1, where the 
discrepancies are clear. We also plot the original $M\propto \sigma^4$
relation (\ref{msignum}) for comparison.

\begin{figure}
\includegraphics[width=\columnwidth]{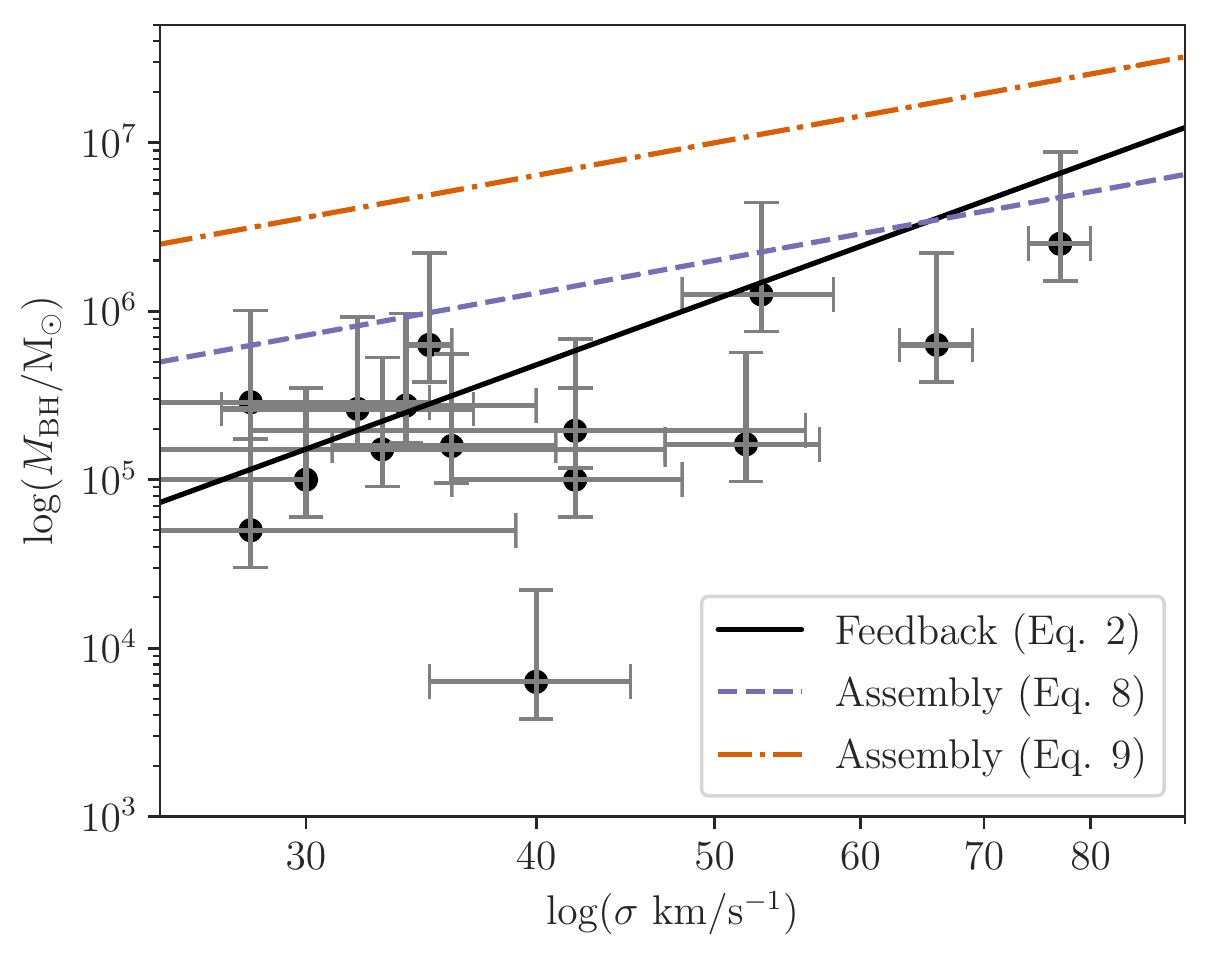}
\caption{$M$-$\sigma$ relation using the data from Baldassare et al. 2020 and the references quoted therein. The $M \propto \sigma^2$ relations (\ref{msig1}, \ref{msig2}) predicted by the assembly picture  are the orange (dashed) and blue (dash--dot) lines, while the black (solid) curve is the original 
$M \propto \sigma^4$ relation (\ref{msignum}) predicted by feedback.
}
\end{figure}

\section{The Black--Hole vs Bulge--Mass Relation for Dwarf Galaxies}

The Faber--Jackson--like relation (\ref{fjdwarf}) for dwarfs implies that their total 
stellar masses 
vary as $\sigma^2$ rather than $\sigma^4$. Then accepting that
$M_{\sigma} \propto\sigma^4$ as in the sample studied by Baldassare et 
al. (2019), means that we no longer
get a linear relation like (\ref{scaling}) between $M$ and 
$M_b$. Eliminating $\sigma$ 
between eqns (\ref{msignum}, \ref{fj}) instead gives
\begin{equation}
M \simeq 4\times 10^{4}M_9^2\msun R_{\rm kpc}
\label{mm2}
\end{equation}
if the SMBH masses are close to $M_{\sigma}$. We include a 
factor $R_{\rm kpc}= R_b/(1\, {\rm kpc})$ 
(the near--constant radius of low--mass galaxies
predicted by  (\ref{fjdwarf}, \ref{iso})) 
to allow for enforcing continuity between the 
high--mass and low--mass FJ relations (\ref{fj1}, \ref{fjdwarf}) at slightly 
different $\sigma$.

Fig. 2 compares the best--fit value of 
(\ref{mm2}) with the AGN sample of Baldassare et al. (2020) 
Table 1 and Davis et al., 2020), 
with the $M - M_b$ relation found by Schutte, 
Reines \& Greene (2019) plotted for comparison. This figure
suggests that SMBH
are less massive relative to their hosts at low galaxy masses, perhaps because the
stellar feedback fixing $M_b$ is less effective in removing gas before it makes 
stars. Garratt--Smithson et al., (2019) suggest that this does happen, because
gradual stellar feedback delays the unbinding of most of the gas. Instead it makes
`chimneys' in the dense shell surrounding the hot feedback region, 
venting the hot gas from the galaxy before it can remove much of the 
star--forming gas.

%
%



\begin{figure}
\includegraphics[width=\columnwidth]{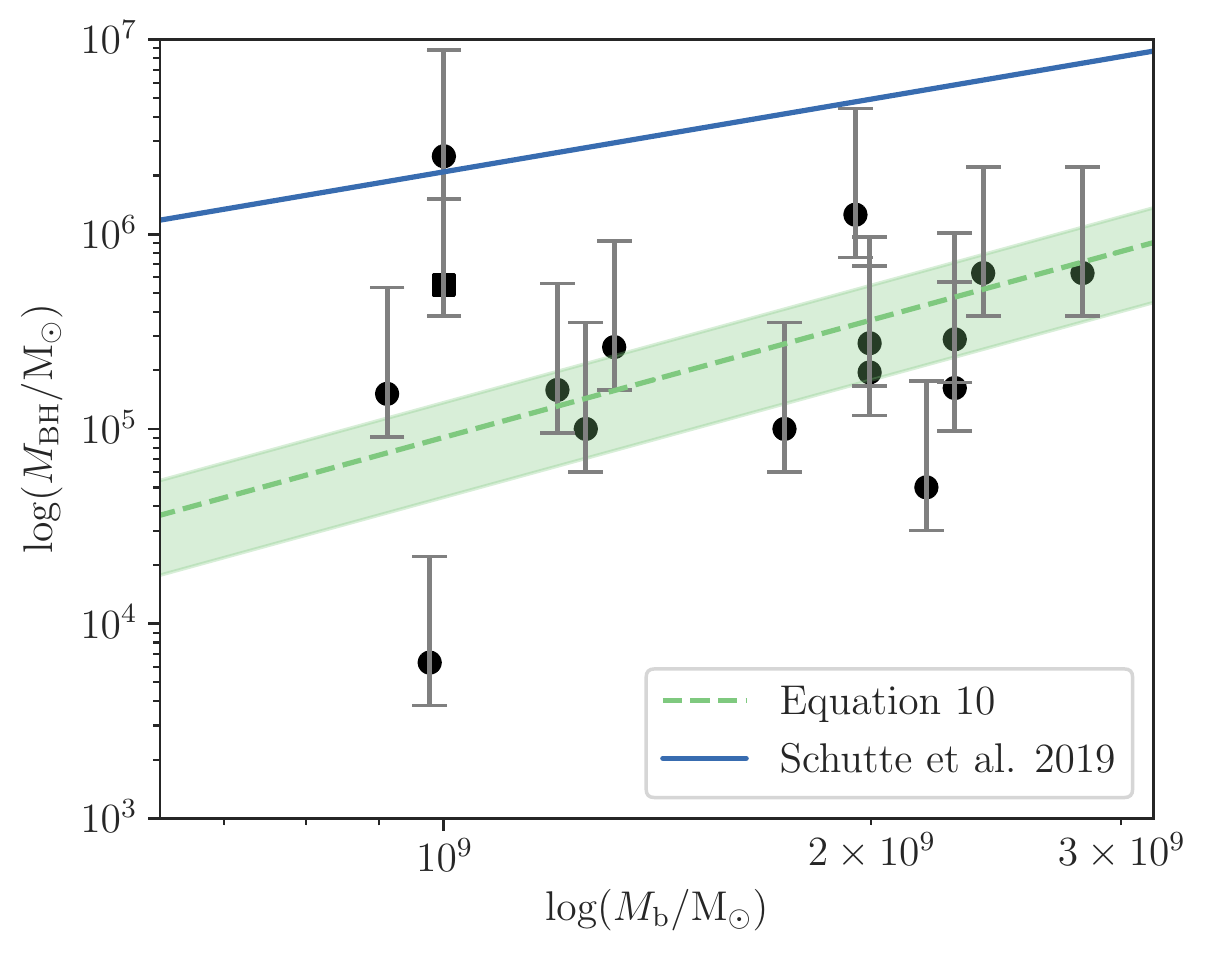}
\caption{The quadratic $M - M_b$ 
relation (\ref{mm2}, with best--fit value $R_b = 2.3 \pm 1.1$\, kpc) 
for low--mass galaxies plotted against the data from 
Table 1 of Baldassare et al (2020) and references therein, together with 
the point from Davis et al., (2020) (as a square). 
The best--fit 
linear $M - M_b$ relation found by Schutte, Reines \& Greene (2019) for 
the full SMBH sample for all galaxies is plotted for comparison. This used
photometric modeling and colour--dependent mass--to--light ratios to determine 
$M_b$. }
\end{figure}

It appears that the central black holes in dwarf galaxies play a similar active role 
in their evolution as in more massive galaxies, even though they may be
relatively less massive compared with their hosts. 
L\"asker et al., (2016) and Nguyen et al. (2019b)
also find black hole masses lying below a linear 
extrapolation of the $M - M_b$ scaling relation in dwarf galaxies.
This presumably makes 
them harder to discover, supporting the arguments of Kaviraj et al. (2019) for a 
large black hole occupation fraction in dwarfs. This would probably require
even smaller initial seed SMBH masses $M$ in the assembly picture than the 
decrease of a factor 100 we estimated in Section 2, while the fundamental 
difficulty in fitting the observed $M - \sigma$ relation would remain unchanged.

Of course dwarf galaxies are
not a homogenous group, and in particular there is likely to be a sub--population 
where the central black hole has not grown to an energetically significant mass $\sim M_{\sigma}$ (cf Pacucci et al., 2018, King \& Nealon, 2019). 

\section{Outflows from dwarf galaxies}

A second recent set of observations offers another test of the origin of the scaling 
relations. Manzano--King et al. (2019: hereafter MK19) give
spatially--resolved kinematic measurements of AGN--driven outflows in dwarf 
galaxies in the stellar mass range $M_b\sim 6\times 10^8 - 9\times 10^9\msun$
These are selected from SDSS DR7 and DR8 and followed up with Keck/LRIS 
spectroscopy. In a total sample of 50 dwarf galaxies,
they find ionized gas 
outflows out to distances up to $1.5\,{\rm kpc}$ in 13, all having velocities 
above the escape value for their dark
matter halos. There are line--ratio indications of AGN activity in 9 of the 13 
galaxies with outflows, and in
6 of these the outflow appears to be driven by the AGN rather than a starburst, 
with one further less clear example. 

Although mild outflows are allowable in the assembly picture, they
have no particular significance there. But powerful
outflows are an inevitable and tightly constrained consequence of the feedback picture (cf King, 
2003; King 2005; Zubovas \& King 2012, summarized in  King \& Pounds, 
2015). Once $M$ reaches the value (\ref{msignum}), 
all of the mechanical energy of the UFO wind 
is communicated to the host's bulge 
ISM in a forward shock, driving this gas 
away in an energy--driven outflow. In an isothermal potential this has speed
\begin{equation}
v_{\rm out} \simeq 
1230\sigma_{200}^{2/3}\left(\frac{lf_c}{f_g}\right)^{1/3}\,{\rm km\,s^{-1}}
\label{vout}
\end{equation}
(King 2005, Zubovas \& King, 2012).
Here $l \sim 1$ is the ratio of the driving SMBH accretion luminosity to the Eddington value,
and $f_c \simeq 0.16$ is the cosmological mean value of $f_g$. (The 
dark matter halo at larger radii is irrelevant for 
the baryonic physics determining $v_{\rm out}$ and  $M_{\sigma}$.)
The corresponding mass outflow rate is
\begin{equation}
\dot M_{\rm out} = 3700\sigma_{200}^{8/3}l^{1/3}\, \msun\, {\rm yr}^{-1}
\label{mout}
\end{equation}
where $f_g$ has been taken equal to $f_c = 0.16$ (in King \& Pounds (2015)
the corresponding equation [(57)] gives the exponent of $\sigma$ incorrectly as 
10/3 rather than 8/3). Once the energy--driven 
outflow described by (\ref{vout}, \ref{mout}) begins to escape the 
baryonic part of the galaxy 
it accelerates above
the speed (\ref{vout}) (cf Zubovas \& King 2012b).


There is a large body of observational data (cf references in the review of King 
\& Pounds, 2015, Section 5.3) showing that many massive
galaxies drive out gas roughly as described by (\ref{vout}, \ref{mout}). 
In applying this formalism to dwarf galaxies we in principle need
velocity dispersions $\sigma$. These are not measured
for the MK19 sample, but
inspection of Fig. 2 of MK19 confirms that (\ref{r}) is a reasonable
approximation for the visible size of these galaxies. Then we use (\ref{fj}) to 
eliminate $\sigma$ from eqn (\ref{vout}) in favour of $M_b$. (This procedure 
also avoids the need to estimate the stellar mass--to--light ratio.) We find
\begin{equation}
v_{\rm out} \simeq 307M_9^{1/3}x^{1/3}\, {\rm km\,s^{-1}}
\label{voutnum}
\end{equation}
where  $M_9 = M_b/10^9\msun$, and
\begin{equation}
x = \frac{lf_c}{R_{\rm kpc}f_g}
\label{x}
\end{equation}
with $R_{\rm kpc} = R_b/{\rm kpc}\simeq 1$ the radius of the visible galaxy. 
We expect $x$ to have similar values $\sim 0.5$ 
for all 8 dwarfs with AGN--driven outflows in the sample of MK19. MK19 do not measure
black hole masses $M$, but these do not appear in the expressions (\ref{vout}, \ref{mout}) as we have assumed that $M$ has reached the limiting value (\ref{msignum}) and triggered an 
energy--driven outflow.

\begin{figure}
\includegraphics[width=\columnwidth]{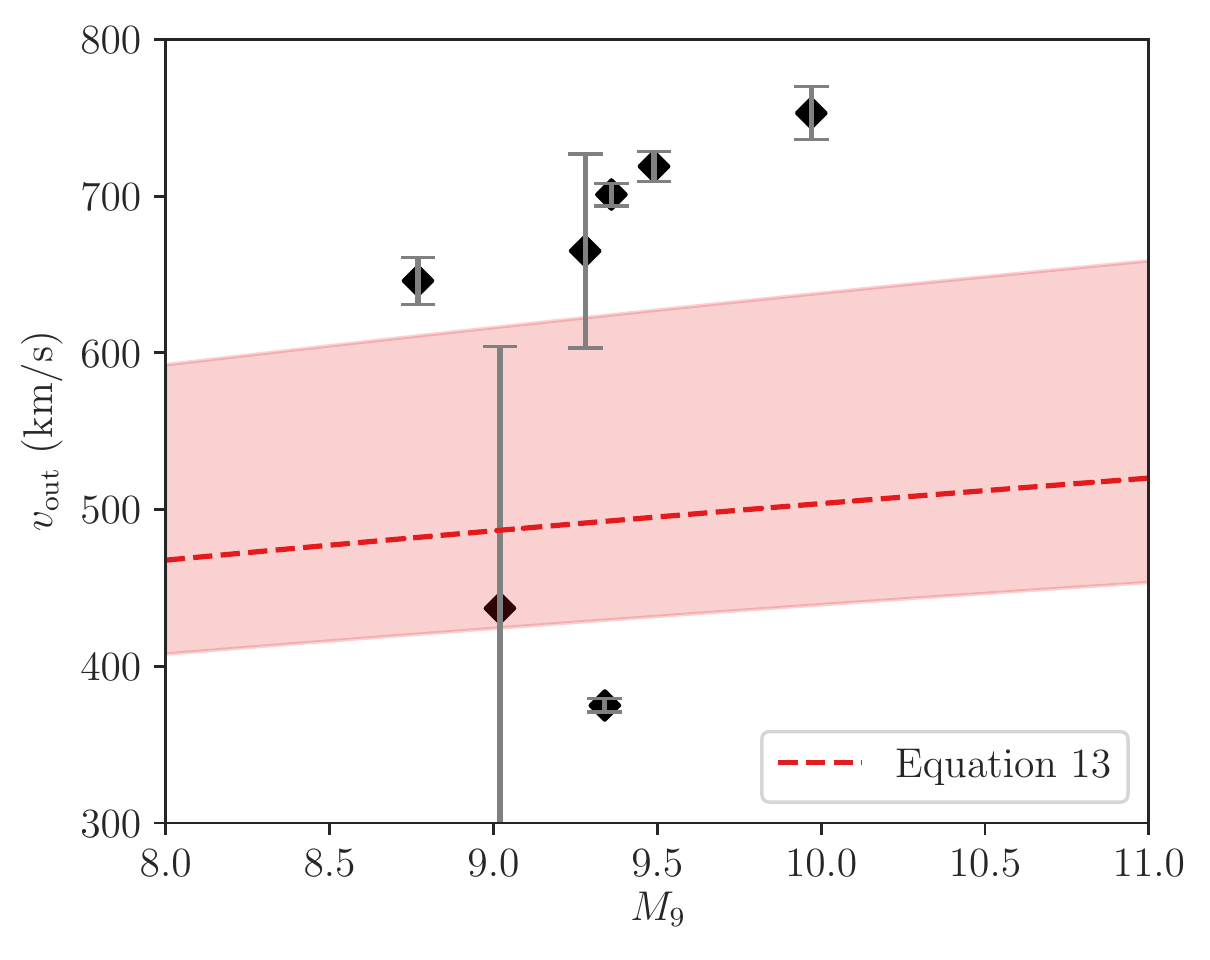}
\caption{The outflow velocity $v_{\rm out}$ (column 7 of Table 1 of MK19)
versus galaxy mass $M_9$ in the dwarf AGN sample of MK19. In almost all cases these are significantly higher than would be predicted (equation \ref{voutnum}, shown) for outflows which had not yet escaped the visible galaxy. Here we use $R_{\rm kpc}$ fit in Figure 2, with the uncertainties shown in the shaded region. The exception is
J084234.51+031930.7, the only one whose narrow lines give
a composite BPT classification. }
\end{figure}

Fig. 3 compares the data of MK19 with (\ref{voutnum}), using the fitted value of $R_{\rm kpc}$ from Fig. 2 (corresponding to $x = 0.44 \pm 0.30$). It is immediately obvious from Fig. 3 that all but one 
of the observed outflow velocities are significantly larger than given by 
(\ref{voutnum}), as we expect if the outflows have already largely escaped 
the visible galaxies. This is strongly suggested by their large spatial scales, of order 
the half--light radii. (The exception is J084234.51+031930.7, 
the only one whose narrow lines give a composite BPT classification.)

Similarly we expect that the mass outflow rate in these galaxies should currently 
be somewhat higher than the
values $\dot M_{\rm out} \simeq 100\msun\,{\rm yr}^{-1}$ predicted by 
(\ref{mout}) with 
$\sigma \sim 50\,{\rm km\, s^{-1}}$. Then if feedback is continuous, a galaxy 
would lose most of its gas 
in a total time 
\begin{equation}
t_{\rm deplete}\sim \frac{f_gM_b}{\dot M_{\rm out}} \sim 10^6 
\frac{M_9}{\sigma_{50}^{8/3}}\, {\rm yr}
\end{equation}
where $M_9 = M_b/10^9\msun$, and we have used (\ref{iso}, \ref{r}). Again
replacing $\sigma_{50}$ with $M_9$ from (\ref{fj}) we find
\be
 t_{\rm deplete}\sim 2.5\times 10^6M_9^{-1/3}\,{\rm yr}.
\ee
So we estimate depletion times of a few million years for all dwarf galaxies 
in the energy--driven stage of AGN feedback expected as the SMBH mass approaches $M_{\sigma}$, only weakly dependent on galaxy mass.

\section{Conclusion}
 
Recent observations extend the tight $M - \sigma$ relation found for massive 
galaxies to dwarf galaxies with low--mass ($M\sim 10^5 - 10^6\msun$) black 
holes. This is
natural if feedback causes the scaling relations, but hard to reconcile with
the assembly picture. The initial $(M, M_b)$ seed pairs must be much smaller, 
and have a far tighter dispersion than thought. Independently of
these significant adjustments, assembly always gives a linear $M, M_b$
relation. Then the empirical Faber--Jackson--like relation (\ref{fj}) for 
dwarf galaxies means that the assembly picture predicts a significant flattening
in the slope of the $M-\sigma$ relation for black holes in dwarf galaxies, from
$M \sim \sigma^4$ to $M\sim \sigma^2$. These predicted higher--mass 
SMBH are not found, even though all selection effects would favour this.
There is no degree of freedom in the assembly picture
to overcome this problem, as the assumption of a 
linear $M - M_b$ relation arising from the central limit theorem is fundamental 
to it.

The apparent inadequacy of the assembly picture in explaining the 
SMBH--galaxy relations at all masses arises because it implicitly assumes that the
relation between $M$ and $M_b$ drives the $M - \sigma$ relation. It appears
instead that the fundamental physical scaling relation is between $M$ and $\sigma$, and is caused by SMBH feedback, as already strongly suggested by the 
wide discrepancy between SMBH and bulge binding energies. The 
$M - M_b$ relation between black hole and
 bulge mass is acausal, arising from the quite 
independent connection between $M_b$ and $\sigma$ set by stellar rather than 
black--hole feedback.

The discovery of powerful AGN--driven winds rapidly
removing gas from dwarf galaxies gives additional support to the feedback picture. This is in line with recent
work (cf de Nicola, Marconi \& Longo, 2019; Chen et al., 2020) on massive 
galaxies. Further observations of dwarf galaxies and their central black holes are
likely to give critical insights into the origins of the black hole -- galaxy scaling
relations.


\section*{Acknowledgments}

We thank Joop Schaye for very helpful comments on cosmological simulations,
and the referee for a perceptive report.
This project has received funding from the European Research Council (ERC) 
under the 
European Union's Horizon 2020 research and innovation programme (grant 
agreement No 681601). RN acknowledges support from UKRI/EPSRC through a Stephen Hawking Fellowship (EP/T017287/1).

\section*{Data Availability Statement}
No new data were generated or analysed in support of this research.

{}

\end{document}